\documentstyle[amstex,righttag,12pt,verbatim]{article}
\newtheorem{th}{Theorem}

\numberwithin{equation}{section}

\newcommand{\D}{\operatorname{d}}

\newcommand{\I}{\operatorname{i}}
\begin{document}
\title{The  Three-Wave  Resonant Interaction:\\
Deformation of the Plane Wave Solutions\\ and
Darboux Transformations}  
\author{Francisco Guil and
Manuel Ma\~nas\thanks{M.\ M.\ acknowledges  partial 
support from  CICYT  proyecto PB92--019} 
\\
Departamento de F\'\i sica Te\'orica\thanks{FAX \#: 34 1 394 51 97}, 
Universidad Complutense\\ E28040-Madrid, Spain\\
manuel@@eucmos.fis.ucm.es}

\date{}

\maketitle


\begin{abstract}
The plane wave solutions of the three-wave resonant interaction
in the plane   are considered. It is shown that rank-one constraints over
the right derivatives of invertible operators on an arbitrary linear space
gives  solutions of the three-wave resonant interaction
that can be understood as a Darboux transformation of the 
plane wave solutions.
The method is extended further to obtain 
general Darboux transformations: for any solution of the three-wave
interaction problem and vector solutions of the corresponding
Lax pair large families of new solutions,
expressed in terms of Grammian type determinants of these vector
solutions, are given.
\end{abstract}

\newpage

\section{Introduction}

This section is devoted to the description of  what we call
the 3-wave resonant interaction (3WRI) system, that
contains under reduction the 
3-wave resonant interaction equations.
 This motivates our choice for the name, nevertheless
notice that it can be considered as the 3-component
Kadomtsev-Petviasvilii equations.

We shall consider six complex amplitudes
 $\{p_{ij}\}\begin{Sb}i,j=1,2,3\\i\neq j\end{Sb}$
depending on three complex variables $z_1,z_2,z_3$.
The 3WRI system is the following set of equations,
\begin{equation}\label{3wris}
\partial_kp_{ij}+p_{ik}p_{kj}=0\text{, for distinct $i,j,k=1,2,3$},
\end{equation}
where $\partial_i:=\partial/\partial z_i$.
These equations have a Lax pair
first derived in \cite{zs} and given in characteristic coordinates
in \cite{ah}. In fact there are two of such linear systems, one
adjoint to the other. Given three functions $F_i$, $i=1,2,3$,
 Eqs. (\ref{3wris}) are the compatibility conditions
for the  linear system 
\begin{equation}
  \label{lp}
  \partial_j F_i+p_{ij}F_j=0, \text{  for $i\neq j$},
\end{equation}
or for the adjoint linear system for $\tilde F_i$, $i=1,2,3$,
\begin{equation}
  \label{alp}
  \partial_j \tilde F_i+p_{ji}\tilde F_j=0, \text{  for $i\neq j$}.
\end{equation}

The 3WRI equations appear when we require $p_{ij}=p_{ji}^*$ and $\text{Im}z_i=0$
so that $z_i=x_i\in\Bbb R$.
Using  the notation $q_k:=p_{ij}$, $i,j,k$ cyclic,
 the Eqs. (\ref{3wris}) become
\begin{equation}
\label{3wri}
\partial_kq_k+q_i^*q_j^*=0,
\end{equation}
the well known 3WRI equations. 
These equations are relevant in the context of fluid dynamics, nonlinear
optical phenomena \cite{abdp} and plasma physics \cite{crs}, but see 
\cite{zmr} and \cite{krb} as well. 
The inverse scattering problem associated to the Lax pair,
as given in \cite{ah},
was studied in \cite{c}, but it was 
in the series of papers \cite{k,k1} where the direct and inverse
scattering problem was solved.
The 3WRI equations also posses infinite-dimensional symmetry
algebras \cite{lms,mw}, B\"acklund transformations \cite{k2,lps}
and have interesting reductions to the Painlev\'e equations \cite{lms,mw,k}.

Eqs. (\ref{3wris}) have two obvious symmetries. Firstly, we consider
changes in the amplitude. The action in the moduli space is
 \begin{equation}
\label{s1}
p_{ij}(z_1,z_2,z_3)\mapsto\exp(a_i(z_i)-a_j(z_j))p_{ij}(z_1,z_2,z_3)
\end{equation} 
for arbitrary functions $a_i(z_i)$, $i=1,2,3$. Another symmetry
 is a scaling transformation defined by 
any set of non-zero complex numbers 
$\{s_{ij}\}\begin{Sb}i,j=1,2,3\\i\neq j\end{Sb}\subset\Bbb C^{\times}$
with $s_{12}s_{23}s_{31}=s_{13}s_{32}s_{21}$ as 
\begin{equation}
\label{s2}
p_{ij}(z_1,z_2,z_3)\mapsto s_{ij}p_{ij}(S_1z_1,S_2z_2,S_3z_3),
\end{equation}
 with $S_k=s_{ik}s_{kj}/s_{ij}=s_{jk}s_{ki}/s_{ji}$, with $i,j,k$ cyclic,
that provides an action on the solution space.

In this paper a special r\^ole will be played
by the exponential solutions  or
 plane wave solutions of the 3WRI equations. Let us consider
under which conditions the trial functions
\[
p_{ij}(z_1,z_2,z_3)=\lambda_{ij}\exp(\sum_{m=1}^3a_{ijm}z_m)
\]
are solutions of the Eqs. (\ref{3wris}).
One can easily check that we need  that the following conditions  hold:
\begin{align}
a_{ijm}&=a_{ikm}+a_{kjm}, \label{a}\\
\lambda_{ij}a_{ijk}&=\lambda_{ik}\lambda_{kj}\label{l}
\end{align}
with  $i,j,k,m=1,2,3$  and $i,j,k$ distinct.
From Eqs. (\ref{a}) it follows that $a_{ijm}+a_{jim}=0$, and once these
conditions are fulfilled we only need to request $a_{12m}+a_{23m}+a_{31m}=0$.
We write $a_{iji}=:-\mu_{ij}$, 
so that $a_{ijj}=-a_{jij}=\mu_{ji}$ and by (\ref{a}) (for $m=k$) 
$a_{ijk}=a_{ikk}+a_{kjk}=\mu_{ki}-\mu_{kj}$. In fact, this is the more
general parametrization of the solutions of Eqs. (\ref{a}). Therefore,
taking $\lambda_{ij}$ and $\mu_{ij}$ subject to
\begin{equation}
\label{lm}
\lambda_{ij}(\mu_{ki}-\mu_{kj})=\lambda_{ik}\lambda_{kj},
\end{equation}
for $i,j,k=1,2,3$ and different, the more general plane wave solution
of Eqs. (\ref{3wris}) is given by
\begin{equation}
\label{vacuum}
p_{ij}^{(0)}(z_1,z_2,z_3)=\lambda_{ij}\exp(-\sum_{k\neq i}z_k\mu_{ki}+
\sum_{k\neq j} z_k\mu_{kj}).
\end{equation}

 Observe that there is a compatibility condition over the
amplitudes $\lambda_{ij}$, namely
\begin{equation}
\label{l3}
\lambda_{12}\lambda_{23}\lambda_{31}+\lambda_{13}\lambda_{32}\lambda_{21}=0,
\end{equation}
and if the $\lambda$'s do not vanish this equation  itself gives
the possible $\lambda$'s that when plugged into Eq. (\ref{lm}) give
the differences $\mu_{ki}-\mu_{kj}$. That only the differences are fixed
is a consequence of  the symmetry of the 3WRI system defined in
(\ref{s1}). Indeed, given a plane wave with parameters
 $\{\lambda_{ij},\mu_{ij}\}$  then 
the set $\{\lambda_{ij},\mu_{ij}+a_i\}$ defines another possible plane
wave
(here we have taken the functions
$a_i(z_i)=a_iz_i$). Observe also that Eqs. (\ref{lm}) are invariant under
the substitution 
$\{\lambda_{ij},\mu_{ij}\}\mapsto \{s_{ij}\lambda_{ij},S_i\mu_{ij}\}$,
 a consequence of the symmetry transformation (\ref{s2}).

 The two symmetries given by (\ref{s1}) and (\ref{s2}) of the
3WRI system remain symmetries of Eqs. (\ref{3wri}) when
$\text{Re}a_i=0$, and $a_i(x_i)$ takes imaginary values,
$s_{ij}=s_{ji}^*\in\Bbb C^\times$ and $S_k\in\Bbb R^{\times}$. For the plane wave solutions of (\ref{3wri})
we need $\lambda_{ij}=\lambda_{ji}^*$ and also $a_{ijm}=a_{jim}^*$,
but $a_{jim}=-a_{ijm}$ and therefore $a_{ijm}\in \I \Bbb R$. That is,
the $\mu_{ij}$ are imaginary numbers, and the $p_{ij}$ are really
physical plane waves, i. e. with no damping. Define $a_{ijm}=:\I k_{ijm}$
and $\mu_{ij}=:\I m_{ij}$, with $k_{ijm}, m_{ij}\in\Bbb R$,
we have: $k_{iji}=m_{ij}, k_{ijj}=-m_{ji}$ and
$k_{ijk}=m_{ik}-m_{jk}$. 
We also introduce the notation $\lambda_{ij}=\ell_k$, $i,j,k$ cyclic.
Then, the plane wave solutions are
\[
q_k(x_1,x_2,x_3):=\ell_k\exp(\I\sum_{m=1}^3k_{ijm}x_m),
\]
where the amplitudes $\ell_i$ and the wave vectors defined by the $m$'s
satisfy
\[
\I \ell_k(m_{ki}-m_{kj})=\ell_i^*\ell_j^*
\]
with the indices $i,j$ and $k$  cyclic.

The motivation of this paper comes
from our previous work, \cite{gma,gm1,gm2,gm3}. The main idea in it is to 
consider rank-one constraints on the right derivatives of
certain invertible operators. This was done in \cite{gma} for the
Kadomtsev-Petviashvilii equation and extended to the Davey-Stewartson
equations in \cite{gm1,gm3}, in \cite{gm2} we studied a deformation
of the dromion solution of DSI arising naturally from our method.
Section 2 is devoted to the study of this rank-one constraints,
that in this case are connected with the plane wave solutions of
(\ref{3wris}). Next, in \S 3 we show that the solutions obtained in section
2 generalize to a deformation of the plane wave solutions.
This motivates a further extension in section 4 where a Darboux transformation
for the 3WRI is given. For  any solution we consider vector solutions of the
associated Lax pairs, in terms of which we construct Grammian type determinants
that allow us to give large families of new solutions.

\section{Rank-one constraints and the three-wave resonant interaction}

In this section we shall show how invertible operators can give
solutions to the 3WRI system given by Eqs. (\ref{3wris}).
Consider a function $\psi(z_1,z_2,z_3)$ of the three complex
variables $\{z_1,z_2,z_3\}$  taking values on $\text{GL}(V)$,
the set of invertible operators on some  complex linear space
$V$. On this function we  impose some differential
constraints, namely its right derivatives are of the following form
\begin{equation}
\partial_i\psi\cdot\psi^{-1}=A_i+e_i\otimes\alpha_i,\hspace{1cm}
i=1,2,3\label{frc}
\end{equation}
where $A_i$ are constant operators on $V$, $e_1,e_2,e_3$ are three
independent constant vectors on $V$ and 
$\alpha_i(z_1,z_2,z_3)$ takes values on the set of linear functionals
over $V$, the dual space $V^*$. Now, we must take care of the
compatibility conditions for Eqs. (\ref{frc}). In order to have a set
of closed conditions we require 
\begin{enumerate}
\item[i.] The operators $A_i$ must commute among them.
\item[ii.] The image of the operator $A_i$ when acting on the vector
$e_j$, $i\neq j$, must be expanded by $e_i$ and $e_j$:
\[
A_ie_j=\lambda_{ij}e_i+\mu_{ij}e_j,\,\, i\neq j
\]
where $\{\lambda_{ij},\mu_{ij}\}\begin{Sb}i,j=1,2,3\\ i\neq j\end{Sb}
\subset\Bbb C$ .
\end{enumerate}

The coefficients $\lambda_{ij}$ and $\mu_{ij}$ are not completely
free, indeed there is a further compatibility condition: 
$[A_i,A_j]e_k=0$ for $i,j,k$ different. 
This condition is just
Eq. (\ref{lm}). Concrete realizations of such operators are easily constructed in any space
$V$ although we do not need the explicit form of them.
The compatibility conditions arising from the rank-one
constraints for the right derivatives of  $\psi$ are
\begin{equation}
  \label{alfa}
  (\partial_j-\mu_{ji})\alpha_i+\pi_{ij}\alpha_j+\alpha_iA_j=0,
\end{equation}
where
\[
\pi_{ij}:=\lambda_{ij}+\langle\alpha_i,e_j\rangle.
\]
The contraction of Eq. (\ref{alfa}) with the vector $e_k$, $k\neq i,j$, gives
\[
(\partial_j+\mu_{jk}-\mu_{ji})\pi_{ik}+\pi_{ij}\pi_{jk}=0,
\]
and these equations  can be simplified by defining
\[
p_{ij}:=\exp(-\sum_{k\neq i}z_k\mu_{ki}+\sum_{k\neq
  j}z_k\mu_{kj})\pi_{ij},
\]
to obtain Eqs. (\ref{3wris}) for the functions $p_{ij}$.

Therefore, we have shown how rank-one constraints over the
right derivatives of an invertible operator give rise to solutions
of the 3WRI system what in turns implies that
solving our constrained system  allows
us to find solutions of the 3WRI.

To construct suitable operators $\psi$ we introduce the following
linear functionals on $V$:
\[
\beta_i:=\exp(-\sum_{k\neq i}z_k\mu_{ki})\alpha_i\psi,
\]
 so that
$(\partial_i-A_i)\psi=\exp(\sum_{k\neq i}z_k\mu_{ki})e_i\otimes\beta_i$
 and the compatibility
conditions $[\partial_i-A_i,\partial_j-A_j]\psi=0$ read
\begin{equation}
  \label{beta}
  \partial_j\beta_i+p^{(0)}_{ij}\beta_j=0,\,\, i\neq j,
\end{equation}
with $p_{ij}^{(0)}$ as given in Eq. (\ref{vacuum}).

We also introduce $\psi_0:=\exp(\sum_i z_iA_i)$, $\varphi:=\psi_0^{-1}\cdot
\psi$ and 
\[
b_i:=\exp(\sum_{k\neq i}z_k\mu_{ki})\psi_0^{-1}e_i.
\]
 Then, the rank-one conditions (\ref{frc})
 on the right derivatives of $\psi$ determine that 
\begin{equation}
\label{phi}
\partial_i\varphi=b_i\otimes\beta_i.
\end{equation}

Conversely, given operators $A_i$ as prescribed before 
and the related objects
$\psi_0,b_i$, as well as solutions $\beta_i$ to Eq. (\ref{beta})
we can integrate Eq. (\ref{phi}) and then obtain $\psi=\psi_0\cdot\varphi$
as required.

Summarizing, we can construct solutions of the 3WRI system as follows:  

Given three commuting operators $A_1,A_2,A_3$ on a complex linear
space $V$, three independent linear vectors $e_1,e_2,e_3$, such that
\[
A_ie_j=\lambda_{ij}e_i+\mu_{ij}e_j, \,\, i\neq j
\]
for $i\neq j$, where the elements of $\{\lambda_{ij},\mu_{ij}\}\begin{Sb}i,j=1,2,3\\
i\neq j\end{Sb}\subset\Bbb C$ satisfy:
\[
(\mu_{ki}-\mu_{kj})\lambda_{ij}=\lambda_{ik}\lambda_{kj},
\]
with $i,j,k=1,2,3$ different,
we define the three vector functions 
\[
b_i=\exp(\sum_{k\neq i}z_k\mu_{ki})\psi_0^{-1}e_i,\,\, i=1,2,3
\]
where $\psi_0=\exp(\sum_i z_i A_i)$ and the three linear functionals
$\beta_i$, $i=1,2,3$, subject to
\[
 \partial_j\beta_i+p^{(0)}_{ij}\beta_j=0, i\neq j,
\]
with $p^{(0)}_{ij}=\lambda_{ij}\exp(-\sum_{k\neq i}z_k\mu_{ki}+
\sum_{k\neq j}z_k\mu_{kj})$. If we define
 an invertible operator $\varphi$ by
 the compatible equations
\[
\partial_i\varphi=b_i\otimes\beta_i. 
\]
 then the functions 
\begin{equation}\label{pij}
p_{ij}:=p^{(0)}_{ij}+\langle\beta_i,\varphi^{-1}b_j\rangle
\end{equation}
solve
\[
\partial_jp_{ik}+p_{ij}p_{jk}=0,
\]
the 3WRI system.

Notice the different r\^ole played by the $b$'s
and the $\beta$'s. The $\beta_i$ are simply solutions of Eq. (\ref{beta})
while the definition
 of the $b_i$ is given in terms of
the $A_i$ and the vectors $e_i$.
Nevertheless, both need of the coefficients $\{\lambda,\mu\}$  defined by
Eqs. (\ref{lm}). 
However, one can show that in fact the $b$'s do satisfy analogous equations to
those defining the $\beta$'s, namely
\begin{equation}
  \label{b}
  \partial_jb_i+p^{(0)}_{ji}b_j=0,
\end{equation}
which can be considered adjoint to (\ref{beta}).

We can also seek  wave functions solving the Lax pair or its
adjoint for the solutions given in the previous theorem:

The functions
\begin{align*}
F_i&=\beta_i\varphi^{-1}\\
\tilde F_i&=\varphi^{-1} b_i,
\end{align*}
satisfy Eqs. (\ref{lp}) and Eqs. (\ref{alp}) respectively, where
the expression for the
amplitudes $p_{ij}$ is given in (\ref{pij}).

The proof is just a simple check. First take the derivative with respect to $z_j$
of $F_i$
\[
\partial_j F_i=(\partial_j\beta_i)\varphi^{-1}-
F_i(\partial_j\varphi\cdot\varphi^{-1}),
\]
then use Eqs. (\ref{beta}) and (\ref{phi}) to evaluate the derivatives
of $\beta_i$ and $\varphi$ and to obtain Eq. (\ref{lp}) with
$p_{ij}$ defined in (\ref{pij}).
For $\tilde F_i$ we proceed in an analogous manner:
 \[
\partial_j \tilde F_i=-(\varphi^{-1}\cdot\partial_j\varphi)
\tilde F_i+\varphi^{-1}\partial_jb_i,
\]
but now we need equations (\ref{b}) and (\ref{phi}). 

\section{Deformations of the plane waves for the three-wave resonant interaction}

Equations (\ref{b}) do not characterize the $b$'s, but one can easily
show that in order to construct solutions of the 3WRI system we only
need solutions of the linear Eqs. (\ref{b}) and (\ref{beta}). 
Suppose  that we have $b_i$, $i=1,2,3$, 
vector functions satisfying Eqs. (\ref{b}),
$\beta_i$, $i=1,2,3$, linear functionals  that are solutions of 
Eqs. ({\ref{beta}), $\varphi$ a solution of (\ref{phi}) and define $p_{ij}$ as in (\ref{pij}).
Then, we can evaluate the derivative of $p_{ij}$ with respect to $z_k$ to get:
\begin{multline*}
\partial_kp_{ij}=(-\mu_{ki}+\mu_{kj})\lambda_{ij}\exp(-\sum_{k\neq i}z_k\mu_{ki}
+\sum_{k\neq j} z_k\mu_{kj})\\ +
\langle\partial_k\beta_i,\varphi^{-1}b_j\rangle-
\langle\beta_i,\varphi^{-1}(\partial_k\varphi)\varphi^{-1}b_j\rangle+
\langle\beta_i,\varphi^{-1}\partial_k b_j\rangle
\end{multline*}
and using Eqs. (\ref{lm}), (\ref{beta}), (\ref{b}) and (\ref{phi}) we find out
\begin{align*}
\partial_kp_{ij}&=-p^{(0)}_{ik}p^{(0)}_{kj}-
p^{(0)}_{ik}\langle\beta_k,\varphi^{-1}b_j\rangle-
p^{(0)}_{kj}\langle\beta_i,\varphi^{-1}b_k\rangle-
\langle\beta_i,\varphi^{-1}b_k\rangle
\langle\beta_k,\varphi^{-1}b_j\rangle\\ &=-p_{ik}p_{kj},
\end{align*}
as desired.

Moreover, we can construct wave functions and its adjoints as before.
Summarizing,

\begin{enumerate}
\item[{\bf i.}]
 Given  $\{\lambda_{ij},\mu_{ij}\}\begin{Sb}i,j=1,2,3\\ i\neq j\end{Sb}\subset\Bbb C$ 
 subject to
\[
(\mu_{ki}-\mu_{kj})\lambda_{ij}=\lambda_{ik}\lambda_{kj}\text{, $i,j,k=1,2,3$
 and distinct}
\]
the plane wave solutions of the 3WRI system (\ref{3wris}) are
\[
p^{(0)}_{ij}=\lambda_{ij}\exp(-\sum_{k\neq i}z_k\mu_{ki}+
\sum_{k\neq j} z_k\mu_{kj}).
\]

\item[{\bf ii.}]
Deformations of the plane wave solutions are constructed as follows:
Take  vector functions $b_i(z_1,z_2,z_3)\in V$, $i=1,2,3$,
where $V$ is a complex linear space,
 solutions of:
\[
\partial_jb_i+p_{ji}^{(0)}b_j=0,
 \] 
define linear functionals $\beta_i(z_1,z_2,z_3)\in V^*$, $i=1,2,3$,
subject to
\[
\partial_j\beta_i+p_{ij}^{(0)}\beta_j=0, i\neq j,
\]
and integrate the compatible equations
\[
\partial_i\varphi=b_i\otimes\beta_i.
\]
Then the set of functions 
\[
p_{ij}:=p_{ij}^{(0)}+\langle\beta_i,\varphi^{-1}b_j\rangle
\]
solves
\[
\partial_jp_{ik}+p_{ij}p_{jk}=0,
\]
the 3WRI system.

The functions $F_i=\beta_i\varphi^{-1}$, which are  $V^*$-valued,
and $\tilde F_i=\varphi^{-1}b_i$, $V$-valued functions, satisfy
the linear equations
\begin{align*}
\partial_jF_i+p_{ij}F_j&=0,\\
\partial_j\tilde F_i+p_{ji} \tilde F_j&=0,
\end{align*}
the Lax pairs, (\ref{lp}) and (\ref{alp}) respectively.
\end{enumerate}

The solution shown in part {\bf ii.} 
is  a deformation of the plane wave solutions $p_{ij}^{(0)}$ of 
{\bf i.}  because they are a particular case of $p_{ij}$
when
$b_i=0$ and $\beta_i=0$, $i=1,2,3$. 
Thus, these plane wave solutions can be considered as the starting 
solutions we dress in terms of which we obtain
the families of solutions described
above, and hence as our vacuum solutions.

For the 3WRI equation the 
plane waves solutions, that play the r\^ole of
vacuum solutions have the explicit form
\[
q_k^{(0)}=\ell_k\exp(\I(-\sum_{l\neq i}x_lm_{li}+
\sum_{l\neq j}x_lm_{lj})),
\]
$i,j,k$ cyclic,
and can be dressed using the prescription $\beta_i=b_i^\dagger H$
with a linear operator $H$. Here we assume that
$V$ is a pre-Hilbert space (i.e. $V$ has an inner product). 
Observe that Eqs. (\ref{b}) imply that
$b_i^\dagger H$ solves (\ref{beta}).
Now, if at some point $x_i^{(0)}$ we have
 $H^\dagger\varphi(x^{(0)})=
\varphi^\dagger(x^{(0)})H$ then the equations (\ref{phi})
imply that they hold everywhere. With this equality at hand
it is easy to conclude that our prescription gives the desired reduction. 
Thus, we have obtained

\begin{th}\label{t3}
Given complex numbers $\ell_k$, $k=1,2,3$,
and real numbers $m_{ij}$, $i,j=1,2,3$, $i\neq j$,
 subject to the relations
\[
\I \ell_k(m_{ki}-m_{kj})=\ell_i^*\ell_j^*,
\]
for $i,j,k=1,2,3$  cyclic,
take  vector functions $b_i(x_1,x_2,x_3)\in V$, $i=1,2,3$, in
a pre-Hilbert space $V$,
  solutions of
\begin{align*}
\partial_kb_i+q^{(0)}_jb_k&=0\\
\partial_ib_k+(q^{(0)}_j)^*b_i&=0,
\end{align*}
with $i,j,k$ cyclic and the functions $q_k^{(0)}$ given by
\[
q_k^{(0)}=\ell_k\exp(\I(-\sum_{l\neq i}x_lm_{li}+
\sum_{l\neq j}x_lm_{lj})).
\]
If we define an invertible operator
$\varphi$ by the compatible equations
\[
\partial_i\varphi=b_i\otimes b_i^{\dagger}H,
\]
where $H$ is a linear operator in $V$ and  $\varphi$ is chosen
such that at some  $x^{(0)}\in\Bbb R^3$
satisfies $H^\dagger\varphi(x^{(0)})=\varphi(x^{(0)})^\dagger H$, then
\[
q_k:=q_k^{(0)}+b_i^{\dagger}H\varphi^{-1}b_j
\]
with $i,j,k$ cyclic, solves
\[
\partial_kq_k+q_i^*q_j^*=0,
\]
which are the 3WRI equations.

The functions $\tilde F_i=\varphi^{-1}b_i$ satisfy
the linear equations
\begin{align*}
\partial_k\tilde F_i+q_j\tilde F_k&=0\\
\partial_i\tilde F_k+q_j^*\tilde F_i&=0,
\end{align*}
with $i,j,k$ cyclic.
\end{th}

The lump solutions were obtained in \cite{k2}
by means of the one dimensional version  of the Darboux
transformation (a B\"acklund
transformation) written in the next section although
the first of them was previously
found in \cite{z}. These one-lump solutions
correspond to the dressing of the trivial solution $q^{(0)}=0$ 
in the $V=\Bbb C$ case.  The two-lump solutions constructed in that paper
appear in our scheme for $q^{(0)}=0$ and $V=\Bbb C^2$, and the functions $g_i$ and 
$h_i$ of Kaup correspond to  $b_i=(g_i,h_i)^t$.  Observe
that the linear matrix $H$ is chosen as  the identity and
that the
initial condition $\varphi(x^{(0)})$ is what contains free parameters
of the solution. But, this choice has the disadvantage of loosing some
limiting cases in the solution space.
The dressing of a plane wave instead of a zero solution allows us to obtain
more general solutions than the lumps exhibiting
a non-trivial asymptotic behaviour.

We see that $F_i$ satisfy equations (3) of \cite{k}
associated to the direct scattering problem.
 Theorem \ref{t3} could be understood as a Darboux transformation:
We take the plane wave solutions $q_i^{(0)}$
of the 3WRI equations and their corresponding
scattering data given by the wave functions $b_i$, and then apply
the Darboux
transformation  $b_i\mapsto \tilde F_i=\varphi^{-1}b_i$ and
$q_k^{(0)}\mapsto q_k+ b_i^\dagger H\varphi^{-1}b_j$.
 In fact this argument leads to a more general result,
a Darboux transformation for the 3WRI system that will be treated in the
following section.

\section{Darboux transformations for the three-wave resonant interaction}

For a given solution $p_{ij}$ of  Eqs. (\ref{3wris}) 
and a complex linear space $V$
we consider solutions $b_i(z_1,z_2,z_3)\in V$, $i=1,2,3$,
and $\beta_i(z_1,z_2,z_3)\in V^*$, $i=1,2,3$,
of
\begin{align*}
\partial_jb_i+p_{ji}b_j&=0,\\
\partial_j\beta_i+p_{ij}\beta_j&=0.
\end{align*}
By virtue of the previous linear systems
 the following equation holds
\[
\partial_j(b_i\otimes\beta_i)=\partial_i(b_j\otimes\beta_j).
\]
This implies the existence of a local potential, say $\varphi$,
such that
\[
\partial_i\varphi=b_i\otimes\beta_i.
\]
As the operator $\varphi$ is defined up to a constant we suppose that
it can be chosen to be invertible, $\varphi(z_1,z_2,z_3)\in\text{GL}(V)$.
With this operator we construct new  functions $\hat b_i$ and $\hat\beta_i$
as follows
\[
\hat b_i:=\varphi^{-1}b_i\text{ and }
\hat \beta_i:=\beta_i\varphi^{-1},\,\, i=1,2,3.
\]
If we define now
\[
\hat p_{ij}:=p_{ij}+\langle\beta_i,\hat b_j\rangle=
p_{ij}+\langle\hat\beta_i,b_j\rangle=p_{ij}+\langle\beta_i,\varphi^{-1}
b_j\rangle=p_{ij}+\langle\hat\beta_i,\varphi \hat b_j\rangle,
\]
we immediately see that
\begin{align*}
\partial_j\hat b_i+\hat p_{ji}\hat b_j&=0,\\
\partial_j\hat \beta_i+\hat p_{ij}\hat \beta_j&=0.
\end{align*}
so that $\hat p_{ij}$ is a solution again

\begin{th}\label{t4}
Let  $p_{ij}$ be a solution of  Eqs. (\ref{3wris}) and
define $b_i$, $\beta_i$ as solutions of
the linear systems
\begin{align*}
\partial_jb_i+p_{ji}b_j&=0,\\
\partial_j\beta_i+p_{ij}\beta_j&=0,
\end{align*}
with $b_i$, $i=1,2,3$ taking values in some complex linear space
and $\beta_i$, $i=1,2,3$, in its dual.
If $\varphi$ is an invertible solution of the compatible equations
\[
\partial_i\varphi=b_i\otimes\beta_i,
\]
then
\[
\hat p_{ij}=p_{ij}+\langle\beta_i,\varphi^{-1}b_j\rangle
\]
is another solution of Eqs. (\ref{3wris}).
\end{th}
{\bf Proof:}
The result follows from the considerations previous to the theorem.
Nevertheless, a direct check is easy:
\begin{align*}
\partial_k\hat p_{ij}=&\partial_k p_{ij}+\langle
\partial_k\beta_i,\varphi^{-1}b_j\rangle -\langle
\beta_i,\varphi^{-1}(\partial_k\varphi)\varphi^{-1}b_j\rangle+\langle
\beta_i,\varphi^{-1}\partial_kb_j\rangle\\
=&-p_{ik}p_{kj}-p_{ik}(\hat p_{kj}-p_{kj})-(\hat
p_{ik}-p_{ik})(\hat p_{kj}-p_{kj})-
p_{kj}(\hat p_{ik}- p_{ik})\\
=&-\hat p_{ik}\hat p_{kj}.\Box
\end{align*}

This theorem allows us to deform a given solution by  solving the
associated linear problem. We see that the solutions are expressed in
terms of Grammian determinants of the $b$'s and $\beta$'s. The function
$\varphi$ can be expressed as
\[
\varphi(z_1,z_2,z_3)=\int_{\gamma}(\sum_{i=1,2,3}\D z_i\,b_i\otimes\beta_i)
\]
where $\gamma$ is an adequate
path in $\Bbb C^3$ with end point $z_1,z_2,z_3$, such that $\varphi$
has a non-vanishing determinant and $\tau=\det\varphi$ is
the principal tau function.
If we define the operators
$\varphi_{ij}:=\varphi+b_j\otimes(\beta_i-\delta_i\varphi)$,
with $\delta_i(z_1,z_2,z_3)\in V^*$ such that
$\langle\delta_i,b_j\rangle=\delta_{ij}$, and denote
their determinants by $\tau_{ij}=\det\varphi_{ij}$, the associated
tau functions,
we arrive at the expressions $p_{ij}=\tau_{ij}/\tau$.

The reduction to the Eqs. (\ref{3wri}) is compatible with this  Darboux transformation. 
Notice first that the initial solution  $q_k$ of the (\ref{3wri})
can be considered as a solution $p_{ij}$ of Eqs. (\ref{3wris}) subject
to $p_{ij}=p_{ji}^*$.
This reduction condition can be characterized
as follows: given  a solution $b_i$ of
$\partial_jb_i+p_{ji}b_j=0$, with values in a pre-Hilbert
space $V$ and where $z_i\in\Bbb R$,
 and  a linear operator $H$  the linear functional
$\beta_i=b_i^\dagger H$, with values in $V^*$,
satisfies $\partial_j\beta_i+p_{ij}\beta_j=0$ if and only if $p_{ij}=p_{ji}^*$.
For such $b_i$ and $\beta_i$ the associated $\varphi$ solves 
$\partial_i\varphi=b_i\otimes b_i^\dagger H$, so that
$H\varphi^{-1}=(\varphi^{-1})^\dagger H^\dagger$
holds everywhere if it does at  some point, say $x^{(0)}$. 
Over this data we perform the Darboux transformation of Theorem
\ref{t4}.
Thus, $\hat\beta_i=b_i^\dagger H\varphi^{-1}=
 \hat b_i^\dagger\hat H$, with $\hat H=H^\dagger$,
 and $\hat p_{ij}=\hat p_{ji}^*$, as desired.
 
\begin{th}\label{t5}
 Let $q_i$ represent a solution, $i=1,2,3$, of Eqs. (\ref{3wri}) and take solutions
$b_i(x_1,x_2,x_3)$, $i=1,2,3$, with values in  a pre-Hilbert space $V$,
of
\begin{align*}
\partial_k b_i+q_j b_k&=0\\
\partial_i b_k+q_j^* b_i&=0,
\end{align*}
with $i,j,k$ cyclic.
We define the invertible operator $\varphi$ to satisfy
\[
\partial_i\varphi=b_i\otimes b_i^\dagger H,
\]
with a linear operator $H$ and such that
$H^\dagger\varphi(x^{(0)})=\varphi^\dagger(x^{(0)}) H$.
Then,
\[
\hat q_k=q_k+ b_i^\dagger H\varphi^{-1} b_j,
\]
with $i,j,k$ cyclic, is a new solution of Eqs. (\ref{3wri}).
\end{th}

This theorem allows us to deform solutions of the 3WRI equations in terms
of vector solutions of the associated scattering problem. The new
solutions are constructed as special Grammian determinants.
In the  one dimensional case, $V=\Bbb C$, this transformation was
considered in \cite{k2}, see also \cite{lps}. 

\end{document}